\documentstyle[12pt,epsfig]{article}
\newcommand{\Frac}[2]%
{{\textstyle \frac{\mbox{\footnotesize $#1$}\rule[-0.9mm]{0mm}{1mm}}%
{\mbox{\footnotesize $#2$}\rule{0mm}{3.1mm}}}}
\renewcommand{\thefootnote}{\fnsymbol{footnote}}

\newcommand{\footfrac}[2]%

\begin{document}

\begin{titlepage}
\vspace*{-12 mm}
\noindent
\begin{flushright}
\begin{tabular}{l@{}}
\end{tabular}
\end{flushright}
\vskip 12 mm
\begin{center}
{\large \bf
The nucleon's octet axial-charge $g_A^{(8)}$ with chiral corrections
}
\\[14 mm]
{\bf Steven D. Bass}
\\[5mm]
{\em
Institute for Theoretical Physics, \\
Universit\"at Innsbruck,
Technikerstrasse 25, Innsbruck, A 6020 Austria
\\[10mm]
}

{\bf Anthony W. Thomas}
\\[5mm]
{\em
CSSM, School of Chemistry and Physics, \\
University of Adelaide, Adelaide SA 5005, Australia
\\[5mm]
}
\vspace{0.5cm}
\end{center}
\vskip 20 mm
\begin{abstract}
\noindent 
The value of the nucleon's flavour-singlet axial-charge extracted from 
polarised deep inelastic scattering
is sensitive to the value of the octet axial-charge 
$g_A^{(8)}$ which is usually taken from 
an analysis of hyperon $\beta$-decays within the 
framework of SU(3) symmetry, namely $0.58 \pm 0.03$. 
Using the Cloudy Bag model 
we find that the value of $g_A^{(8)}$ is reduced by as much as 20\% 
below the usual phenomenological value. This increases the value of 
the flavour singlet axial charge ($g_A^{(0)}|_{\rm inv}$) 
derived from deep inelastic data and 
significantly reduces the difference between it and $g_A^{(8)}$.
\end{abstract}

\vspace{9.0cm}

\end{titlepage}
\renewcommand{\labelenumi}{(\alph{enumi})}
\renewcommand{\labelenumii}{(\roman{enumii})}
\renewcommand{\thefootnote}{\arabic{footnote}}
\newpage
\baselineskip=6truemm

\section{Introduction}
Polarised deep inelastic scattering experiments have revealed a small 
value for the nucleon's flavour-singlet axial-charge
$g_A^{(0)}|_{\rm pDIS} \sim 0.3$
suggesting that 
the quarks' intrinsic spin contributes little of the proton's spin.
The challenge to understand the spin structure of the 
proton~\cite{bassrmp,bassbook,mpla,Thomas:2008bd,Thomas:2008ga,reya}
has inspired a vast programme of theoretical activity and new experiments.
Why is the quark spin content $g_A^{(0)}|_{\rm pDIS}$ so small ?
How is the spin ${1 \over 2}$ of the proton built up from 
the spin and orbital angular momentum of the quarks and gluons inside ?

The analysis which leads to $g_A^{(0)}|_{\rm pDIS}$ uses the value of 
the nucleon's octet axial-charge 
$g_A^{(8)}$
which is commonly extracted 
from a 2 parameter fit to hyperon $\beta$-decays using SU(3):
$g_A^{(8)} = 0.58 \pm 0.03$ \cite{fec}.
What separates the values of the octet and singlet axial-charges ?
In this paper we examine the chiral corrections to $g_A^{(8)}$.
We base our analysis on the Cloudy Bag model \cite{schreiber,Kazuo}
which has the attractive feature 
that when pion cloud and quark mass effects are turned off the model 
reproduces the SU(3) analysis.
We find that chiral corrections significantly reduce the value of 
$g_A^{(8)}$. This, in turn, has the effect of increasing the value 
of $g_A^{(0)}|_{\rm pDIS}$ and consequently reducing the absolute value of 
the ``polarised strangeness''
extracted from inclusive polarised deep inelastic scattering.

We start by recalling the $g_1$ spin sum-rules, which  
are derived from the dispersion relation for polarised
photon-nucleon scattering and, for deep inelastic scattering,
the light-cone operator product expansion.
One finds that the first moment of  the $g_1$ structure function
is related
to the scale-invariant axial charges of the target nucleon by
\begin{eqnarray}
\int_0^1 dx \ g_1^p (x,Q^2)
&=&
\Biggl( {1 \over 12} g_A^{(3)} + {1 \over 36} g_A^{(8)} \Biggr)
\Bigl\{1 + \sum_{\ell\geq 1} c_{{\rm NS} \ell\,}
\alpha_s^{\ell}(Q)\Bigr\}
\nonumber \\
& &
+ {1 \over 9} g_A^{(0)}|_{\rm inv}
\Bigl\{1 + \sum_{\ell\geq 1} c_{{\rm S} \ell\,}
\alpha_s^{\ell}(Q)\Bigr\}  +  {\cal O}({1 \over Q^2})
 + \ \beta_{\infty}
.
\nonumber \\
\label{eqc50}
\end{eqnarray}

Here $g_A^{(3)}$, $g_A^{(8)}$ and $g_A^{(0)}|_{\rm inv}$ are the
isovector, SU(3) octet and scale-invariant  flavour-singlet axial
charges, respectively. The flavour non-singlet $c_{{\rm NS} \ell}$
and singlet $c_{{\rm S} \ell}$ Wilson coefficients are calculable in
$\ell$-loop perturbative QCD \cite{Larin:1997}. 
The term $\beta_{\infty}$
represents a possible leading-twist subtraction constant
from the circle at infinity when one closes the contour in the
complex plane in the dispersion relation \cite{bassrmp}.
If finite, the subtraction constant affects just the first moment.
The first
moment of $g_1$ plus the subtraction constant, if finite, is equal
to the axial-charge contribution. 

In terms of the flavour dependent axial-charges
\begin{equation}
2M s_{\mu} \Delta q =
\langle p,s |
{\overline q} \gamma_{\mu} \gamma_5 q
| p,s \rangle
\label{eqc55}
\end{equation}
the isovector, octet and singlet axial charges are:
\begin{eqnarray}
g_A^{(3)} &=& \Delta u - \Delta d
\nonumber \\
g_A^{(8)} &=& \Delta u + \Delta d - 2 \Delta s
\nonumber \\
g_A^{(0)}|_{\rm inv}/E(\alpha_s) 
\equiv 
g_A^{(0)} 
&=& \Delta u + \Delta d + \Delta s
.
\label{eqc56}
\end{eqnarray}
Here
$
E(\alpha_s) = \exp \int^{\alpha_s}_0 \! d{\tilde \alpha_s}\,
\gamma({\tilde \alpha_s})/\beta({\tilde \alpha_s})
$
is a renormalisation group factor which 
corrects for the (two loop) non-zero anomalous dimension $\gamma(\alpha_s)$
of the singlet axial-vector current~\cite{kodaira},  
$
J_{\mu5} = 
\bar{u}\gamma_\mu\gamma_5u
                  + \bar{d}\gamma_\mu\gamma_5d
                  + \bar{s}\gamma_\mu\gamma_5s 
$ ,
which is close to one and which goes to one in the limit 
$Q^2 \rightarrow \infty$;
$\beta (\alpha_s)$ is the QCD beta function.
The singlet axial charge, $g_A^{(0)}|_{\rm inv}$,
is independent of the renormalisation scale $\mu$
and corresponds
to 
$g_A^{(0)}(Q^2)$ evaluated in the limit $Q^2 \rightarrow \infty$.

If one assumes no twist-two subtraction constant 
($\beta_{\infty} = O(1/Q^2)$)
the axial charge contributions saturate the first moment
at leading twist.
The isovector axial-charge is measured independently in neutron
$\beta$-decays
($g_A^{(3)} = 1.270 \pm 0.003$ \cite{PDG:2004})
and the octet axial charge is commonly taken 
to be the value extracted 
from hyperon $\beta$-decays assuming a 
2-parameter SU(3) fit
($g_A^{(8)} = 0.58 \pm 0.03$ \cite{fec}). The uncertainty quoted 
for $g_A^{(8)}$ has been a matter of some debate. There is 
considerable evidence that SU(3) symmetry may be badly broken 
and some have suggested that the error on $g_A^{(8)}$ should be 
as large as 25\%~\cite{Jaffe:1988up}.

Using the sum rule for the first moment of $g_1$, given in Eq.~(1), 
polarised deep inelastic scattering experiments have been
interpreted in terms of
a small value for the flavour-singlet axial-charge.
Inclusive $g_1$ data with $Q^2 > 1$ GeV$^2$
give~\cite{compassnlo}
\begin{equation}
g_A^{(0)}|_{\rm pDIS, Q^2 \rightarrow \infty}
=
0.33 \pm 0.03 ({\rm stat.}) \pm 0.05 ({\rm syst.})
\end{equation}
-- considerably smaller than the value of $g_A^{(8)}$
quoted above. 

In the naive parton model $g_A^{(0)}|_{\rm pDIS}$ is interpreted 
as the fraction of the proton's spin which is carried by the intrinsic
spin of its quark and antiquark constituents.
When combined with 
$g_A^{(8)} = 0.58 \pm 0.03$ 
this value corresponds to a negative strange-quark polarisation
\begin{equation}
\Delta s_{Q^2 \rightarrow \infty}
=
{1 \over 3}
(g_A^{(0)}|_{\rm pDIS, Q^2 \rightarrow \infty} - g_A^{(8)})
=
- 0.08 \pm 0.01 ({\rm stat.}) \pm 0.02 ({\rm syst.})
\end{equation}
-- that is,
polarised in the opposite direction to the spin of the proton.
New fits have been performed which also include data 
from semi-inclusive polarised 
deep inelastic scattering as well as polarised proton 
proton collisions at RHIC.
De Florian et al.~\cite{vogelsang}
take as input $g_A^{(8)}=0.59 \pm 0.03$
and find values $g_A^{(0)} \sim 0.24$ and 
$\Delta s \sim -0.12$, 
with the ``polarised strangeness'' coming almost 
entirely from small values of $x$ outside the measured 
kinematic region -- i.e., for Bjorken $x$ between 0 and 0.001.

There has been considerable theoretical effort to understand
the flavour-singlet axial-charge in QCD.
QCD theoretical analysis leads 
to the formula~\cite{bassrmp,ar,et,ccm,Bass:1991yx}
\begin{equation}
g_A^{(0)}
=
\biggl(
\sum_q \Delta q - 3 {\alpha_s \over 2 \pi} \Delta g \biggr)_{\rm partons}
+ {\cal C}_{\infty}
.
\label{eqa10}
\end{equation}
Here $\Delta g_{\rm partons}$ is the amount of spin carried
by polarised gluons in the polarised proton
($\alpha_s \Delta g \sim {\rm constant}$ as 
 $Q^2 \rightarrow \infty$~\cite{ar,et})
and
$\Delta q_{\rm partons}$ measures the spin carried by quarks
and
antiquarks
carrying ``soft'' transverse momentum $k_t^2 \sim P^2, m^2$
where
$P$ is a typical gluon virtuality
and
$m$ is the light quark mass.
The polarised gluon term is associated with events in polarised
deep inelastic scattering where the hard photon strikes a
quark or antiquark generated from photon-gluon fusion and
carrying $k_t^2 \sim Q^2$~\cite{ccm,Bass:1991yx}.
${\cal C}_{\infty}$ denotes a potential non-perturbative gluon
topological contribution
which is associated with 
the possible subtraction 
constant in the dispersion relation for $g_1$ 
and Bjorken $x=0$~\cite{bassrmp}:
$g_A^{(0)}|_{\rm pDIS} = g_A^{(0)} - C_{\infty}$.

There is presently a vigorous programme to disentangle the different
contributions involving experiments in semi-inclusive polarised deep
inelastic scattering and polarised proton-proton 
collisions~\cite{mpla,Mallot:2006,compass}. 
These direct measurements show no evidence for negative polarised 
strangeness in 
the region $x> 0.006$ and suggest 
$|- 3 {\alpha_s \over 2 \pi} \Delta g| < 0.06$
corresponding to $|\Delta g| < 0.4$ with $\alpha_s \sim 0.3$.
That is, 
they are not able to account 
for the difference
$( g_A^{(0)}|_{\rm pDIS} - g_A^{(8)} ) = -0.25 (\pm 0.07)$
obtained in the analysis of \cite{compassnlo}, 
or $\sim -0.35$ in \cite{vogelsang}.

\section{SU(3) breaking and $g_A^{(8)}$}
\begin{table}[b!]
\caption{$g_A/g_V$ from $\beta$-decays with $F=0.46$ and $D=0.80$,
together with the mathematical form predicted 
in the MIT Bag with
effective colour-hyperfine interaction (see text and~\cite{hogaasen}).}
\vspace{3ex}
{\begin{tabular}{llllr}
Process             &  measurement      &  SU(3) combination & Fit value
& MIT + OGE
\\
\hline
$n \rightarrow p$     &  $1.270 \pm 0.003$  &  $F+D$   & 1.26 
& ${5 \over 3} B' + G$
\\
$\Lambda^0 \rightarrow p$ & $0.718 \pm 0.015$ & $F+{1 \over 3}D$ & 0.73 
& $B'$
\\
$\Sigma^- \rightarrow n$  & $-0.340 \pm 0.017$ & $F-D$ & -0.34 
& $-{1 \over 3} B' - 2G$
\\
$\Xi^- \rightarrow \Lambda^0$ & $0.25 \pm 0.05$ & $F-{1 \over 3}D$ & 0.19 
& ${1 \over 3} B' - G$
\\
$\Xi^0 \rightarrow \Sigma^+$ & $1.21 \pm 0.05$ & $F+D$ & 1.26 
& ${5 \over 3} B' + G$
\\
\hline
\end{tabular}}
\end{table}

Given that the contributions to $g_A^{(0)}$ from the measured distribution   
$\Delta s$ and from $-3{\alpha_s \over 2 \pi} \Delta g$ 
are small, 
it is worthwhile to ask about the value of $g_A^{(8)}$.
The canonical value of $0.58$ 
is extracted from a 2 parameter fit to hyperon $\beta$-decays 
in terms of the SU(3) constants $F=0.46$ and $D=0.80$~\cite{fec}
-- see Table 1.
The fit is good to $\sim 20\%$ accuracy~\cite{Jaffe:1988up,leaders}.
More sophisticated fits will also include chiral corrections.
Calculations of non-singlet axial-charges in relativistic 
constituent quark models 
are sensitive
to the confinement potential, 
effective colour-hyperfine interaction~\cite{myhrer,Close:1978}, 
pion and kaon clouds 
plus additional wavefunction corrections~\cite{schreiber}.
The latter are often treated phenomenologically and chosen 
to reproduce the physical value of $g_A^{(3)}$.

Here we discuss these effects first in the MIT Bag and then in an 
extended 
Cloudy Bag model calculation~\cite{Kazuo}, where chiral corrections 
are in-built.
We focus on $g_A^{(3)}$ and $g_A^{(8)}$.
The Cloudy Bag was designed to model confinement and spontaneous chiral 
symmetry breaking, taking into account pion physics and the manifest 
breakdown of chiral symmetry at the bag surface in the MIT bag.
If we wish to describe proton spin data including matrix elements of
$J_{\mu 5}^3$, $J_{\mu 5}^8$ and $J_{\mu 5}$,
then we would like to know that the model versions of these currents
satisfy the relevant Ward identities.
For the non-singlet axial-charges $g_A^{(3)}$ and $g_A^{(8)}$, 
corresponding to the matrix elements of partially conserved
currents, the model is well designed to make a solid prediction.
For the singlet axial-charge 
the situation is less clear 
since one has first to make an ansatz about 
the relationship between 
the (partially conserved) semi-classical model current and
the QCD current which includes the QCD axial anomaly, including 
the possible topological contribution~\cite{bassrmp}
\footnote{That is, 
 should the model be describing
 $g_A^{(0)}$ at some low scale or one of the scale independent
 quantities $\Delta q_{\rm partons}$, $E(\alpha_s) g_A^{(0)}$ ?
 How should the topological contribution 
 (if finite) be included in the model current ?
 There are also 
 gauge dependence issues if one extrapolates
 matrix elements of the partially conserved QCD singlet axial-vector 
 current away from the forward direction to look at generalised 
 parton distributions 
 and the spin dependence of deeply virtual Compton scattering
 \cite{bassrmp,Jaffe:1988up,Bass00}.}
.

We start with the MIT Bag model.
There are a number of issues associated with the calculation.
First one must account for the effect of the confining potential; 
then the colour-hyperfine interaction; next one must evaluate the 
corrections associated with spurious centre of mass motion (CM) and 
recoil effects. When we turn to the chiral corrections we shall also 
have to choose the 
chiral representation, in particular whether the original surface
coupling or the later volume coupling~\cite{Thomas:1981ps}
version. Since the latter tends to reproduce the empirical value 
of $g_A^{(3)}$ without any CM or recoil 
corrections~\cite{Kazuo} and those are not 
really on a solid theoretical foundation and therefore treated 
phenomenologically, the choice of whether or not to include CM 
and recoil corrections depends on which chiral representation 
is eventually to be used.

For the MIT Bag, the nucleon matrix element of the axial-vector 
current is~\cite{Donoghue:1975yg}
\begin{eqnarray}
\int d^3 x
\sum_i
\langle p s |
\overline{q}_i \vec{\gamma} \gamma_5 
q_i
| p s \rangle
&=&
\int_{\rm bag}
d^3 x
\psi^{\dagger}_i (x) \gamma_0 \vec{\gamma} \gamma_5 \psi (x)
\nonumber \\
&=&
N^2
\int_0^R dx x^2
\biggl\{ 
j_0^2 ({\omega x \over R}) - {1 \over 3} j_1^2  ({\omega x \over R})
\biggr\}
\nonumber \\
&=&
1 - {1 \over 3} \biggl( {2 \omega - 3 \over \omega - 1} \biggr)
= 0.65
\end{eqnarray}
when we substitute for the MIT Bag wavefunction $\psi(x)$.
(Here $\omega=2.04$, 
 $R$ is the bag radius and $N$ is the wavefunction normalisation.)
This factor
(0.65 for massless quarks, 0.67 for quarks with mass $\sim 10$ MeV)
is the crucial difference from non-relativistic constituent quark models.

The effective colour-hyperfine interaction has the quantum numbers of 
one-gluon exchange (OGE).
In models of hadron spectroscopy this interaction 
plays an important role in the nucleon-$\Delta$ and $\Sigma-\Lambda$ 
mass differences, 
as well as the nucleon magnetic moments~\cite{Close:1978} and the spin 
and flavor dependence of parton distribution functions~\cite{Close:1988br}.
It shifts total angular-momentum between spin and orbital 
contributions and, therefore, also contributes to model calculations of 
the octet axial-charges~\cite{myhrer}.
We denote this contribution,  
which has been evaluated to be 0.0373 \cite{myhrer,ushio}
in the MIT Bag (without centre of mass corrections), as ${\cal G}$.

One also has to include additional wavefunction corrections associated 
with the well known issue that,
for the MIT and Cloudy Bag models,
the nucleon wavefunction is not translationally invariant and the 
centre of mass is not fixed.
Corrections to $g_A^{(3)}$ arising from these effects have
been estimated to be as large 
as 15-20\%~\cite{Donoghue:1979ax,Wong:1980ce}. 
Including such a 
correction on the original MIT prediction for $g_A^{(3)}$  
yields a value which is in excellent agreement with experiment.
To compare the model results with experiment
we take the view~\cite{schreiber}
that, in principle,  
the model - with corrections -
should give the experimental value of $g_A^{(3)}$.
We therefore choose the centre-of-mass factor, $Z_{\rm MIT}$,  
phenomenologically 
to give the experimental value of $g_A^{(3)}$.
This then fixes the parameters of the model and 
allows us to use it to make a model prediction for $g_A^{(8)}$.

The model predictions with the centre of mass correction chosen so that 
the final answer for $g_A^{(3)}$ is normalised to its physical value are
\begin{eqnarray}
g_A^{(3)}|_{\rm MIT} &=& 
\biggl( g_A^{(3)}|_{\rm bare} + {\cal G} \biggr) \times Z_{\rm MIT}
\nonumber \\
g_A^{(8)}|_{\rm MIT} &=& 
\biggl( g_A^{(8)}|_{\rm bare} - 3 {\cal G} \biggr) \times Z_{\rm MIT}
\end{eqnarray}
The step by step MIT Bag calculation is shown in Table 2 for massless
quarks.
Note that at the level of Table 1 without additional physics input, 
e.g. pion chiral corrections,
there is a simple algebraic relation between the SU(3) parameters $F$ and $D$, 
the bag parameter $B^\prime$ and the OGE correction $G$:
\begin{eqnarray}
F &=& {2 \over 3} B' - {1 \over 2} G 
\nonumber \\
D &=& B' + {3 \over 2} G \, .
\end{eqnarray}
Substituting the values 
$F=0.46$ and $D=0.80$ 
gives 
$B'=0.73$ and $G=0.05$.
The values 
$G = Z_{\rm MIT} {\cal G} = 0.042$ 
and
$B' = Z_{\rm MIT} \times 0.65 = 0.73$
are in very good agreement with the values extracted from the SU(3) fit 
to hyperon $\beta$-decays~\cite{hogaasen}.
\begin{table}[t!]
\caption{\label{tab:table2}
MIT Bag model calculation 
}
\vspace{3ex}
\begin{tabular}{lcc}
  & $g_A^{(3)}$ & $S_z$ (singlet axial-charge)
\\
\hline
Non-relativistic         & +1.66 & +1.00 \\
Relativistic factor      & +1.09 & +0.65 \\
+ OGE (${\cal G}$ factor)  & +1.13 & +0.54 \\
+ centre of mass         & +1.27 & +0.61 \\
\hline
\\
\end{tabular}
\end{table}

The pion cloud of the nucleon also renormalises the nucleon's axial
charges by shifting intrinsic spin into orbital 
angular momentum~\cite{myhrer,Thomas:2008bd}.
In the Cloudy Bag Model (CBM)~\cite{Thomas:1984}, 
the nucleon wavefunction is written as a Fock expansion
in terms of a bare MIT nucleon,
$|{\rm N}\rangle$, and baryon-pion, $|{\rm N} \pi \rangle$ and 
$|{\Delta \pi}\rangle$, Fock states.
The expansion converges rapidly and we may safely truncate the Fock 
expansion at the one pion level.
The CBM axial charges are~\cite{schreiber}:
\begin{eqnarray}
g_A^{(3)}
&=&
g_A^{(3)}|_{\rm MIT} \times Z_{\rm CBM} \times
\biggl( 1 - {8 \over 9}  P_{N \pi} - {4 \over 9} P_{\Delta \pi}
            + {8 \over 15} P_{N \Delta \pi} \biggr) 
\nonumber \\
g_A^{(8)}
&=&
g_A^{(8)}|_{\rm MIT} \times Z_{\rm CBM} \times
\biggl(1 - {4 \over 3}  P_{N \pi} + {2 \over 3} P_{\Delta \pi} \biggr) .
\end{eqnarray}
Here,
$Z_{\rm CBM}$ is the phenomenological CM correction factor 
chosen to preserve the physical value of $g_A^{(3)}$ after the 
chiral correction associated with the pion cloud has been included.
The coefficients $P_{N \pi}$ and $P_{\Delta \pi}$
denote the probabilities to find 
the physical nucleon in
the
$|N \pi \rangle$ and $|\Delta \pi \rangle$ 
Fock states,  
respectively and 
$P_{N \Delta \pi}$ is the interference term. There is a wealth of 
phenomenological information to suggest 
that $P_{N \pi}$ is between 20 and 25\%, 
while  $P_{\Delta \pi}$ is in the range 5--10\%~\cite{awtpion}. 
For the interference term we simply follow the calculation of 
Ref.~\cite{schreiber} and set 
$P_{N \Delta \pi}$ = 0.30. If we initially take 
($P_{N \pi}, \, P_{\Delta \pi}$) = (0.20, 0.10),
the bracketed pion cloud renormalisation factors 
in Eq.(10)
are 0.94 for $g_A^{(3)}$ and 0.8 for $g_A^{(8)}$.
With these parameters, the Cloudy Bag prediction 
for the axial-charges is shown in Table 3.
\begin{table}[t!]
\caption{\label{tab:table3}
Bag model calculation with pions included, 
$Z=0.7$, $P_{N \pi} = 0.20$, $P_{\Delta \pi} =0.10$
}
\vspace{3ex}
\begin{tabular}{lcc}
  & $g_A^{(3)}$ & $S_z$ (singlet axial-charge) \\
\hline
Non-relativistic      & +1.66 & +1.00 
\\
Relativistic          & +1.09 & +0.65 
\\
+ OGE                 & +1.13 & +0.54 
\\
+ Pions               & +1.06 & +0.43
\\
+ centre of mass      & +1.27 & +0.52
\\
\hline
\\
\end{tabular}
\end{table}

In order to estimate the model dependent variation, we repeat 
this calculation using the values 
$Z=0.66$, $P_{N \pi}=0.24$, $P_{\Delta \pi} =0.10$
and
$Z=0.70$, $P_{N \pi}=0.24$, $P_{\Delta \pi} =0.06$~\cite{awtpion}.
For these parameter choices, the value of $S_z$ in Table 3 reduces 
to 0.50 and 0.48, respectively.
Thus the bag model, including OGE and pion loop corrections and with
the CM correction adjusted to give the physical value of $g_A^{(3)}$, 
yields a light quark spin content 
$S_z \, = \, 0.50 \pm 0.02$. 
As long as we do not include strange quarks, 
this is also the value of $g_A^{(8)}$ in the model.

However, having found significant effects from the pion cloud, it is 
also reasonable to ask about the effect of the kaon cloud, in particular 
the $K \Lambda$ Fock component of the nucleon wave function. This term, which 
corresponds to a probability of order 5\% or less and naturally explains 
the measured strange electric and 
magnetic form factors of the proton~\cite{Thomas:2005qb,Young:2006jc},  
generates a small $\Delta s$ $\sim -0.01$~\cite{Kazuo}. This would increase 
$g_A^{(8)}$ by 0.02, however, the 
corresponding wave function renormalisation reduces 
the non-strange contribution 
by about 5\%, leaving the combined value of 
$g_A^{(8)} \, \equiv \Delta u + \Delta d - 2 \, \Delta s$ 
with final model prediction between 0.47 and 0.51.
That is, pion and kaon cloud chiral corrections have the potential 
to reduce $g_A^{(8)}$
from the SU(3) value $3F-D$ 
to $0.49 \pm 0.02$, 
which is still within the 20\% variation found in the SU(3) fit in Table 1.
(The corresponding 
semi-classical model value for $g_A^{(0)}$ is $0.46 \pm 0.02$
-- a little higher than reported in Ref.~\cite{myhrer} because of 
our requirement that the same model reproduce the phenomenological 
value of $g_A^{(3)}$.)

\subsection{Volume coupling CBM}

Although the volume coupling version of the CBM was derived by a unitary 
transformation on the original CBM and must therefore be equivalent, it is 
well known that because practical calculations are of necessity carried out 
only to a finite order, it is simpler to understand some physical 
phenomena in one version or the other. In particular, low energy theorems 
such as the Weinberg-Tomozawa relation for s-wave pion scattering are trivial 
to derive within the volume coupling version~\cite{Thomas:1981ps}. 
Within this representation there is an additional correction 
to $g_A^{(3)}$~\cite{Thomas:1981ps}. 
Indeed, in the SU(3) case this correction to the 
pure quark term has the form
\begin{equation}
\delta A^\lambda_i = 
- \frac{1}{2 f_\pi} f_{ijk} \bar{q} 
\gamma^\lambda \lambda_j q \phi^k \theta_V \, ,
\label{eq:deltaA}
\end{equation}
where $f_{ijk}$ are the usual SU(3) structure constants, 
$\phi^k \, , k \in (1,8)$ are the octet Goldstone 
boson fields and $q$ are the quark fields confined in the bag volume $V$.
For $g_A^{(3)}$ this term yields an increase of order 15\%, 
which means that one has essentially no phenomenological 
need for the CM correction in 
order to reproduce the physical value~\cite{Kazuo}. 
On the other hand, this additional 
term does not effect the flavour singlet spin content. 
For $g_A^{(8)}$ the meson loop 
generated by the additional term in Eq.(\ref{eq:deltaA}) 
is only of order 3\%~\cite{Kazuo}, being 
suppressed relative to that for $g_A^{(3)}$ because it 
involves a kaon rather than a pion.  
Thus, in this case the model 
yields values for $S_z$ which more or less correspond to 
the ``+ Pions'' line of Table 3. 
\begin{table}[t!]
\caption{\label{tab:table4}
Volume coupling version of the cloudy bag model, 
including pions and kaons, with the 
pion parameters as in Table 3. Following Ref.~\cite{Kazuo}, 
in the last line $g_A^{(3)}$ has been rescaled to match the 
experimental value and 
$g_A^{(8)}$
and $g_A^{(0)}$ have been rescaled by the same factor.
}
\vspace{3ex}
\begin{tabular}{lccc}
  & $g_A^{(3)}$ & 
$g_A^{(8)}$ 
& $g_A^{(0)}$ \\
\hline
Non-relativistic      & +1.66 & +1.00  &  +1.00
\\
Relativistic          & +1.09 & +0.65  &  +0.65
\\
+ OGE                 & +1.13 & +0.54  &  +0.54
\\
Volume CBM            & +1.29 & +0.45  &  +0.40
\\
rescale               & +1.27 & +0.44  &  +0.39
\\
\hline
\\
\end{tabular}
\end{table}

The full results of the volume coupling CBM calculation 
are summarised in Table 4. Once one allows for the variation in 
the pion-baryon Fock components considered earlier this leads to 
$g_A^{(0)} = 0.37 \pm 0.02$, while $g_A^{(8)} = 0.42 \pm 0.02$. Once again, 
+0.03 of the difference has the same origin, while the correction 
term $\delta A_8$ (c.f. Eq.(\ref{eq:deltaA})) yields the extra +0.02.
\footnote{
These results agree well with the results found in Ref.~\cite{Kazuo},
 namely $g_A^{(8)} = 0.47$ and $g_A^{(0)} = 0.41$}

\section{Concluding Remarks}

We have shown that once one takes care to 
consistently reproduce the experimental 
value of the proton's axial charge, the 
bag model with exchange current corrections 
arising from gluon exchange plus the chiral corrections 
associated primarily with 
the pion cloud lead to a substantial reduction of $g_A^{(8)}$
below the value commonly used in the 
analysis of spin structure functions. 
The extent of the reduction depends upon the 
version of the CBM used, lying in the 
range $0.49 \pm 0.02$ for the original CBM and 
$0.42 \pm 0.02$ for the volume coupling version. 
These changes alone raise the value of 
$g_A^{(0)}|_{\rm pDIS, Q^2 \rightarrow \infty}$ 
derived from the experimental data
from
$0.33 \pm 0.03 ({\rm stat.}) \pm 0.05 ({\rm syst.})$ to 
$0.35 \pm 0.03 ({\rm stat.}) \pm 0.05 ({\rm syst.})$ and 
$0.37 \pm 0.03 ({\rm stat.}) \pm 0.05 ({\rm syst.})$, respectively. 
Both of these values 
have the effect of reducing the level of OZI violation 
associated with the difference 
$g_A^{(0)}|_{\rm pDIS} - g_A^{(8)}$ 
from $0.25 \pm  0.07$ 
to 
just $0.14 \pm 0.06$ 
and $0.05 \pm 0.06$, respectively
\footnote{
 It would be interesting to consider the effect of using the values of 
 $g_A^{(8)}$ calculated here on QCD global fits to polarised 
 deep inelastic as well as proton-proton collision data \cite{vogelsang}
 where the octet axial-charge enters non-linearly.}
.
It is this OZI violation which eventually needs to be explained
in terms of singlet degrees of freedom: 
effects associated with polarised glue and/or a topological effect 
associated with $x=0$.

As we have explained, the remaining uncertainty in this model calculation 
lies in the small 
ambiguity between the two chiral representations that one can choose. 
In order 
to quote an overall value that properly encompasses these possibilities  
we follow the Particle Data Group procedure~\cite{PDG_proc}, 
finding a combined value of $g_A^{(8)} = 0.46 \pm 0.05$ 
(with the corresponding semi-classical
 singlet axial-charge or spin fraction 
being $0.42 \pm 0.07$).
With this final value for 
$g_A^{(8)}$ the corresponding experimental value of $g_A^{(0)}|_{\rm pDIS}$ 
would increase to $g_A^{(0)}|_{\rm pDIS} = 0.36 \pm 0.03 \pm 0.05$.

\section*{\bf Acknowledgements} 
We would like to acknowledge helpful communications from 
F. Myhrer and K. Tsushima.
The research of
SDB is supported by the Austrian Science Fund, FWF, through grant 
P20436, 
while
AWT is supported by the Australian Research Council through an 
Australian Laureate Fellowship and by the University of Adelaide.

\end{document}